\documentclass{aa}
\usepackage[english]{babel}
\usepackage[varg]{txfonts}

\usepackage[usenames,dvipsnames]{xcolor}

\usepackage{epsfig}
\usepackage{graphicx,natbib}
\usepackage{amsmath,amssymb}
\usepackage{hyperref}
\usepackage{wasysym}
\usepackage{natbib}
\usepackage{ulem}
\usepackage{cancel}
\usepackage{subfigure}
\usepackage{gensymb}

\hypersetup{
    colorlinks=true,
    citecolor=blue,
    linkcolor=red,
    urlcolor=black
    }

\def\ga{\,\hbox{\hbox{$ > $}\kern -0.8em \lower 1.0ex\hbox{$\sim$}}\,}
\def\la{\,\hbox{\hbox{$ < $}\kern -0.8em \lower 1.0ex\hbox{$\sim$}}\,}
\def\aap{A\&A}

\begin{document}
\def\nat{Nature }
\def\apj{Astrophys. J. }
\def\apjs{Astrophys. J., Suppl. Ser. }
\def\apjl{Astrophys. J., Lett. }
\def\apss{Astrophys. and Space Science}

\def\gcms{g cm$^2$ s$^{-1}$}
\def\gcc{g cm$^{-3}$}

\title{Impact of the Hall effect in star formation : improving the angular momentum conservation}
\titlerunning{}

\author{P. Marchand \inst{1}, K. Tomida \inst{1,2}, B. Commer\c con \inst{3}, G. Chabrier \inst{3,4}
}

\institute{
Department of Earth and Space Science, Osaka University, Toyonaka, Osaka 560-0043, Japan
\and Department of Astrophysical Sciences, Princeton University, Princeton, NJ 08544
\and CRAL, Ecole normale sup\'erieure de Lyon, UMR CNRS 5574, Universit\'e Claude Bernard Lyon 1, France
\and School of Physics, University of Exeter, Exeter, EX4 4QL, UK}

\authorrunning{P. Marchand et~al.}

\date{}

\abstract{We present here a minor modification of our numerical implementation of the Hall effect for the 2D Riemann solver used in Constrained Transport schemes, as described in \citet{2018A&A...619A..37M}. In the previous work, the tests showed that the angular momentum was not conserved during protostellar collapse simulations, with significant impact. By removing the whistler waves speed from the characteristic speeds of non-magnetic variables in the 1D Riemann solver, we are able to improve the angular momentum conservation in our test-case by one order of magnitude, while keeping the second-order numerical convergence of the scheme. We also reproduce the simulations of \citet{Tsukamoto2015} with consistent resistivities, the three non-ideal MHD effects and initial rotation, and agree with their results. In this case, the violation of angular momentum conservation is negligible in regard to the total angular momentum and the angular momentum of the disk.}


\maketitle

\section{Introduction}

In \citet{2018A&A...619A..37M} (hereafter, paper 1), we presented the numerical implementation of the Hall effect in the AMR code {\ttfamily RAMSES} \citep{teyssier}, aimed for application in protostellar collapse simulations. While the implementation successfully passes several tests, showing the second-order convergence in space, the gas angular momentum is not conserved in star formation simulations. As expected, the Hall effect generates rotation in an initially non-rotating cloud and counter-rotating envelopes form on both sides of the mid-plane \citep{2011ApJ...733...54K,Tsukamoto2015,2017PASJ...69...95T,2017MNRAS.466.1788W}. However, shortly after the formation of the first Larson core, a large amount of rotation is generated in the first core and violates the conservation of the total angular momentum, which is a purely numerical issue. This problem arises in every simulation with the Hall effect, severely limiting the validity of our results. We could not find the origin of the problem. \citet{2011ApJ...733...54K} encountered a similar issue and assumed it was due to their boundary conditions. We however did not find significant angular momentum transport through the box boundaries. In this work, as well as in paper I, we only consider the angular momentum of the fluid, not the magnetic field component, because there is no transfer between both in our framework.

In this paper, we present a minor modification of our numerical scheme that significantly decreases the spurious generation of angular momentum. This method is presented in section \ref{SecMethod}, then its impact on the test case of paper 1 in section \ref{SecNorot}. In section \ref{SecRot}, we make a comparison with a previous study of a more realistic scenario (initial rotation and consistent resistivities), and sections \ref{SecDiscussions} and \ref{SecConclusions} are dedicated to discussion and conclusions.

\section{Methods}\label{SecMethod}

\subsection{The Hall effect implementation}

The implementation of the Hall effect in the eulerian {\ttfamily RAMSES} code, as described in details in paper 1, has been inspired by \citet{Lesur2014}. We briefly summarize it here.

The magnetic field is updated at every time-step using the Constrained Transport scheme \citep{1988ApJ...332..659E} on cell interfaces. For the x-component, the integration reads

\begin{align}
  \frac{B^{n+1}_{x,i-\frac{1}{2},j,k}-B^{n}_{x,i-\frac{1}{2},j,k}}{\Delta t} = &\frac{E^{n+\frac{1}{2}}_{z,i-\frac{1}{2},j+\frac{1}{2},k}-E^{n+\frac{1}{2}}_{z,i-\frac{1}{2},j-\frac{1}{2},k}}{\Delta x} \nonumber \\
                                                                               &- \frac{E^{n+\frac{1}{2}}_{y,i-\frac{1}{2},j,k+\frac{1}{2}}-E^{n+\frac{1}{2}}_{y,i-\frac{1}{2},j,k-\frac{1}{2}}}{\Delta x}.\label{inducfinite}
\end{align}

$B^{n+1}_{x,i-\frac{1}{2},j,k}$ is the x-component of the magnetic field at time-step $n+1$ on cell face $(i-1/2,j,k)$, with $(i,j,k)_{i,j,k \in \mathbb{N}}$ being the indexes of the center of the cell. $E^{n+\frac{1}{2}}_{*,i-\frac{1}{2},*,*}$ are the electric fields on cell edges drawing the contour of the cell interface. These electric fields are computed after the prediction step of the MUSCL scheme \citep{1976cppa.conf...E1V} using the HLL 2D-Riemann solver \citep{2004JCoPh.195...17L}

\begin{align}
  & E^{n+\frac{1}{2}}_{z} = \nonumber \\
  & \frac{S_LS_BE^{n+\frac{1}{2},LB}_{z}-S_LS_TE^{n+\frac{1}{2},LT}_{z}-S_RS_BE^{n+\frac{1}{2},RB}_{z}+S_RS_TE^{n+\frac{1}{2},RT}_{z}}{(S_R-S_L)(S_T-S_B)} \nonumber\\
                                &-\frac{S_TS_B}{S_T-S_B}(B^T_{x} - B^B_{x})+\frac{S_RS_L}{S_R-S_L}(B^R_{y} - B^L_{y}).\label{EqHLL2D}
\end{align}
Here, $E^{n+\frac{1}{2},(LB,LT,RB,RT)}$ represents the electric fields at the corners of the 4 cells adjacent to the edge\footnote{L,B,R,T stand for Left, Bottom, Right, Top}, $B^{(L,B,R,T)}$ are the magnetic fields at the cell corners averaged on the interfaces, and $S_{(L,B,R,T)}$ are the minimum and maximum characteristic wave speeds of the system in the 2 directions perpendicular to the edge. Figure 3 of paper 1 summarizes all the notations.

The electric fields are computed by adding the flux-freezing electric field of ideal MHD $\mathbf{u}\times\mathbf{B}$ and the Hall electric field $\mathbf{u}_\mathrm{H}\times\mathbf{B}$. $\mathbf{u}_\mathrm{H}=-\eta_\mathrm{H}\mathbf{J}/||\mathbf{B}||$ is the Hall speed and is averaged over the 4 cells, $\eta_\mathrm{H}$ is the Hall resistivity and $\mathbf{J}=\nabla \times \mathbf{B}$ is the electric field. Hence, for, e.g.,  corner LB,
\begin{equation}
  E^{n+\frac{1}{2},LB}_z = (u^{n+\frac{1}{2},LB}_x + u_{\mathrm{H},x}) B^{L}_y - (u^{n+\frac{1}{2},LB}_y + u_{\mathrm{H},y}) B^{B}_x.\\
\end{equation}

The very last sentence of section 3.5 in paper 1 states that the whistler waves speeds are accounted for in the 1D-Riemann problems at cell interfaces during the prediction step of the MUSCL scheme. Actually, that sentence is incorrect, as the prediction step does not use the Riemann solvers. The intended meaning is that whistler speeds are added to the wave fan of other variables (density, momentum...), that use a 1D Riemann solver to compute the fluxes at cell interfaces in the correction step. The wave fan should be the same for every variable because the (characteristic) waves speeds correspond to the eigenvalues of the system of equation. The 1D HLL \citep{HLL1983} flux for the non-magnetic variables $U$ reads

\begin{equation}\label{EqHLL1D}
  F_{HLL}^{n+\frac{1}{2}} = \frac{S_R F_L^{n+\frac{1}{2}} - S_L F_R^{n+\frac{1}{2}} + S_L S_R (U_R^{n+\frac{1}{2}}-U_L^{n+\frac{1}{2}})}{S_R S_L},
\end{equation}

with subscripts L and R indicating the left and right side of the interface. The flux is then used in a second-order Godunov scheme to update the flow variables.

The speed of the whistler waves is
\begin{equation}\label{whistlernum}
  c_\mathrm{w} = \frac{\eta_\mathrm{H}\pi}{2\Delta x} + \sqrt{\left(\frac{\eta_\mathrm{H}\pi}{2\Delta x}\right)^2 + c_\mathrm{A}^2},
\end{equation}
with $c_\mathrm{A}=B/\sqrt{\rho}$ the Alfven speed, $\rho$ the density and $\Delta x$ the cell-size. For all variables, the characteristic speeds used in equations \ref{EqHLL2D} and \ref{EqHLL1D} account for the whistler speeds. Since they are usually the fastest waves at high resolution, we have $S_L = u_x-c_\mathrm{w}$, $S_R = u_x+c_\mathrm{w}$, $S_B = u_y-c_\mathrm{w}$ and $S_T = u_y+c_\mathrm{w}$.

\subsection{Modification of the scheme}

The modification we propose consists in not accounting for whistler waves in the Riemann problems of variables other than the magnetic field. In other words, we do not use them to compute $S_L$ and $S_R$ in equation \ref{EqHLL1D}. Two reasons motivate this modification. Firstly, the truncation error increases with the characteristic speeds. At the center of a protostellar collapse simulation, the whistler speed can reach several hundred times the value of the second fastest wave, the fast magneto-sonic wave, which is usually the fastest wave in the absence of the Hall effect. In the test case presented in paper 1, there is a factor $\sim 300$ between both speeds at the end of the protostellar collapse simulation. Truncation errors increase then significantly in the first Larson core even for the purely hydro variable, the momentum in particular. The second reason is that the Hall effect does not directly affect the fluid motion but only indirectly through the Lorentz force (by changing the magnetic fields). 
Removing the whistler speed in the magnetic Riemann problems as well leads to a magnetic field instability.

In appendix \ref{AppConvergence}, we show that our new scheme still propagates the whistler waves at the correct frequencies and keeps its second order convergence in space.

\section{Test case: non-rotating sphere} \label{SecNorot}

\subsection{Initial conditions}

We use the same initial condition and numerical parameters as in section 5 of paper 1, a uniform $1.5$ M$_\odot$ non-rotating sphere of radius $3712$ au at $T=10$ K with a uniform magnetic field $B=90.3$ $\mu$G. The temperature is given by the following equation of state

\begin{equation}\label{eqeos}
  T=10 \left[ 1+\left(\frac{\rho}{10^{-13}~\mathrm{g}~\mathrm{cm}^{-3}} \right)^{\frac{2}{3}} \right] .
\end{equation}

We include only the Hall effect with a constant resistivity $\eta_\mathrm{H}=10^{20}$ cm$^2$ s$^{-1}$. We use the generalized monotonized central (moncen) slope limiter with a coefficient 1.5 as in paper 1, with two different refinement criteria, 8 and 16 points per Jeans length respectively. We also perform one simulation with a moncen coefficient of 1.05 and 8 points per Jeans length for consistency with section \ref{SecRot}.

\subsection{Results}

\begin{figure}
\begin{center}
\includegraphics[trim=5.5cm 2cm 5cm 3.5cm, clip, width=0.45\textwidth]{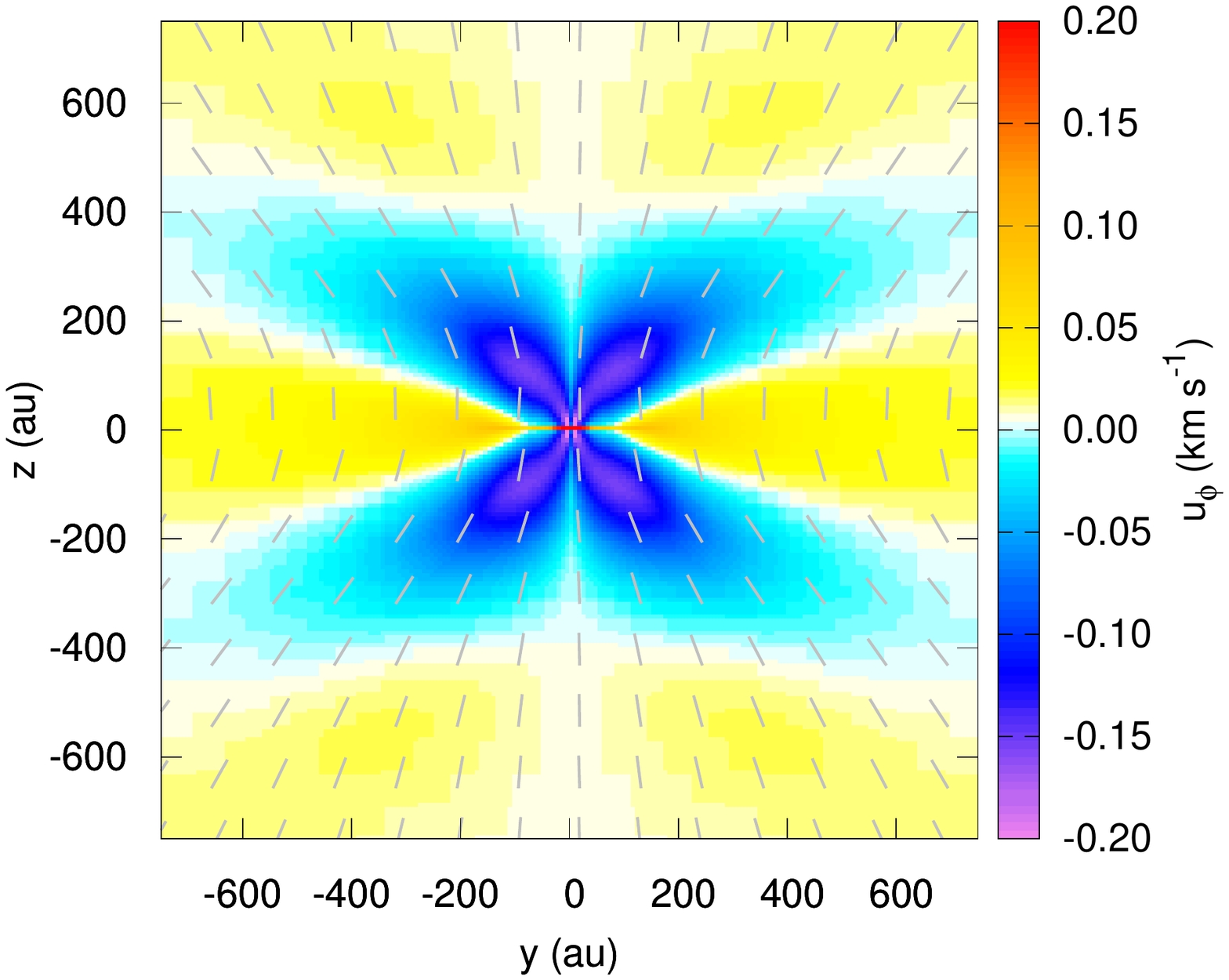}
  \caption{Density slice at $x=0$ of the azimuthal velocity at $t=t_\mathrm{C}+700$ years. Yellow-red colors represent rotation in the positive azimuthal direction, while blue-purple colors represent the negative direction. Grey vectors indicate the direction of the magnetic field.}
  \label{FigVtheta}
\end{center}
\end{figure}

Our results are qualitatively similar to paper 1. The Hall effect generates rotation in the cloud, especially in the mid-plane, and counter-rotating envelopes develop above and below the mid plane to compensate the generation of angular momentum. Figure \ref{FigVtheta} represents a slice at $x=0$ of the azimuthal velocity with magnetic field vectors. The scale and the shape of the counter-rotating envelopes are similar to the collapse in paper 1.

\begin{figure}
\begin{center}
\includegraphics[trim=3cm 1cm 4cm 2cm,width=0.45\textwidth]{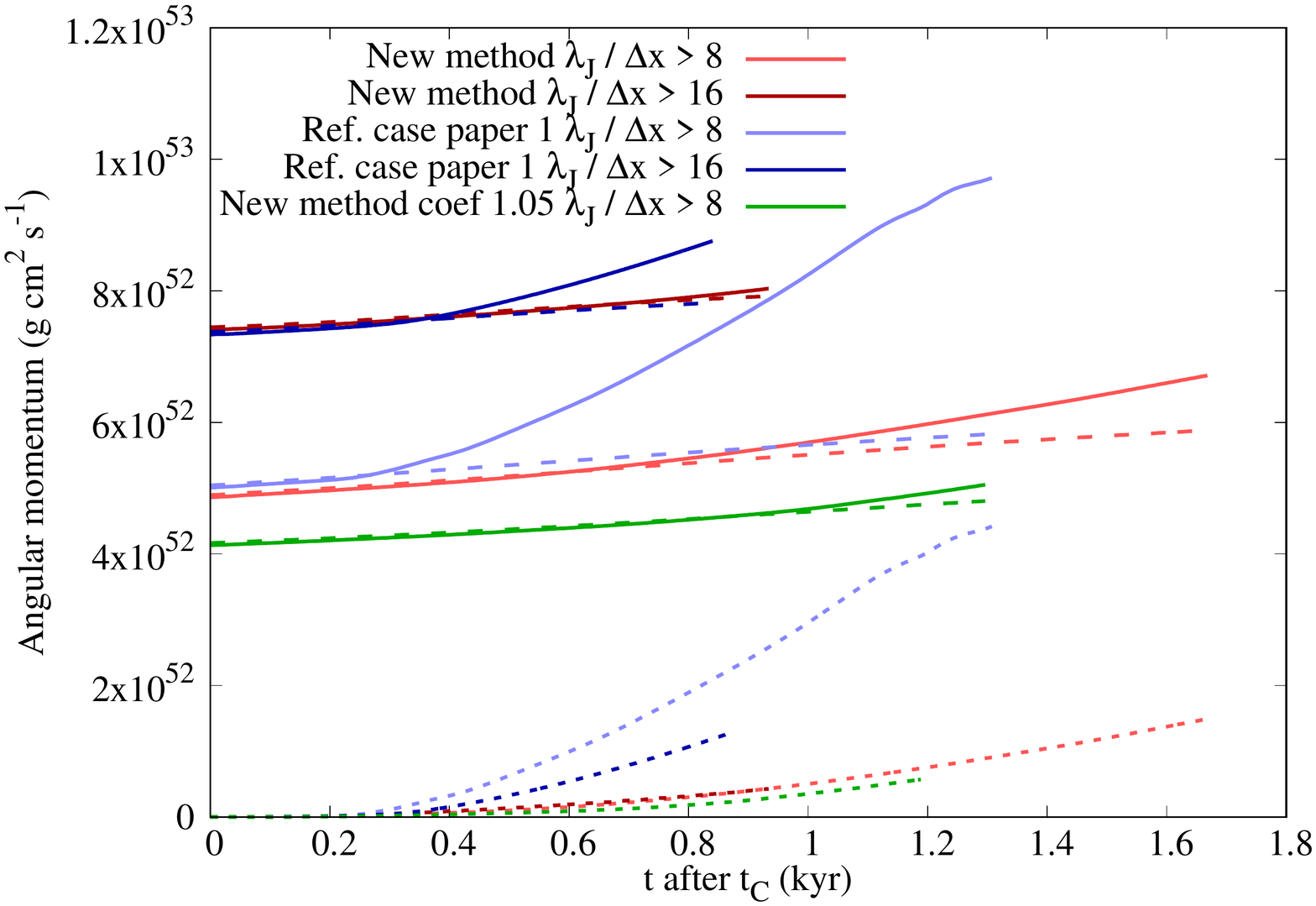}
  \caption{Evolution of the positive and negative angular momentum ($L_z^+$ and $L_z^-$) in solid and dashed line respectively. Dotted lines indicate the angular momentum of the disk. The blue lines represent the reference case of paper 1, while the red and green lines show the evolution for the modified scheme, for a moncen coefficient of 1.5 and 1.05 respectively. The lower resolution simulations with 8 points per Jeans length are in light colors and higher resolution simulations with 16 points per Jeans length are in darker colors.}
  \label{FigMomnorot}
\end{center}
\end{figure}

The "positive" and "negative" angular momenta in the simulations ($L_z^+$ and $L_z^-$ as defined in paper 1) are plotted in figure \ref{FigMomnorot} alongside the reference case of paper 1. There is an obvious improvement of the conservation of the total angular momentum with the present method. The divergence starts at $t_\mathrm{C}+700$ yr instead of $t_\mathrm{C}+300$ yr, and the increase of $L_z^+$ is one order of magnitude slower than in the previous case. The conservation of the lower resolution case is even better than the higher resolution case of paper 1 for both slope limiters. If we compare the excess of angular momentum (difference between solid and dashed lines) to the total angular momentum of the disk (dotted lines), there is also a significant improvement. In the previous method, 85\% of the disk's angular momentum was due to the numerical error. With the new scheme, this fraction decreases to less than 50\%. Additionally, both resolutions and both slope limiters show the same disk's angular momentum.

In this test-case, the initial angular momentum is zero, while in reality dense cores exhibit rotational motions \citep{1993ApJ...406..528G}. Moreover, a realistic Hall resistivity, as computed in \citet{2016A&A...592A..18M}, is likely to be one order of magnitude lower than in this case. The creation of spurious angular momentum would be then damped by this factor approximatively. For these reasons, and the addition of ambipolar and Ohmic diffusion, we can expect the angular momentum increase to be negligible compared to the total angular momentum, and even the accretion of angular momentum in the disk, in a more realistic setup. This point is tested in next section.

\section{Models with initial rotation} \label{SecRot}

\subsection{Initial conditions}

\begin{figure*}
\begin{center}
\includegraphics[trim=7cm 2cm 5cm 3cm, width=0.32\textwidth]{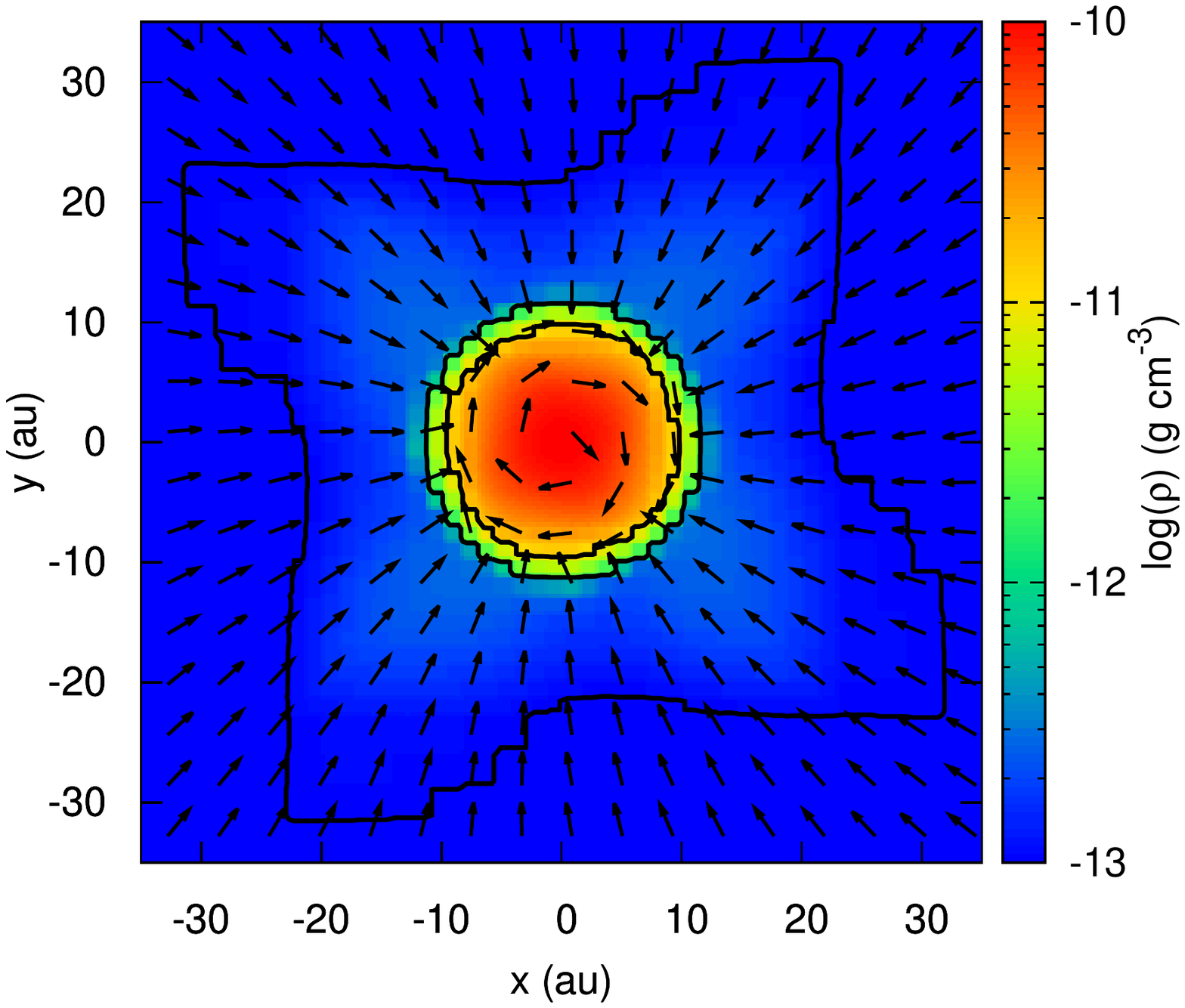}
\includegraphics[trim=7cm 2cm 5cm 3cm, width=0.32\textwidth]{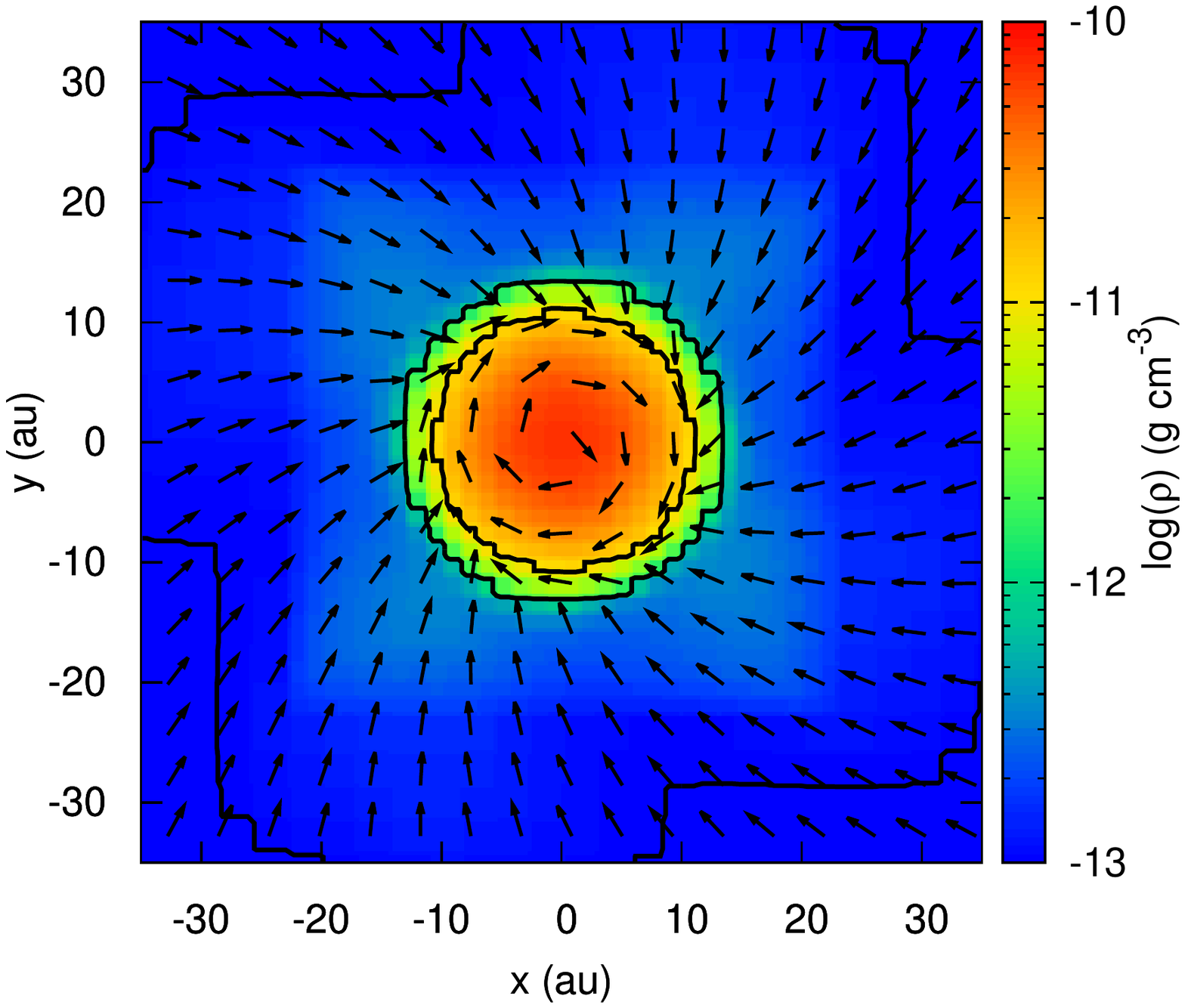}
\includegraphics[trim=7cm 2cm 5cm 3cm, width=0.32\textwidth]{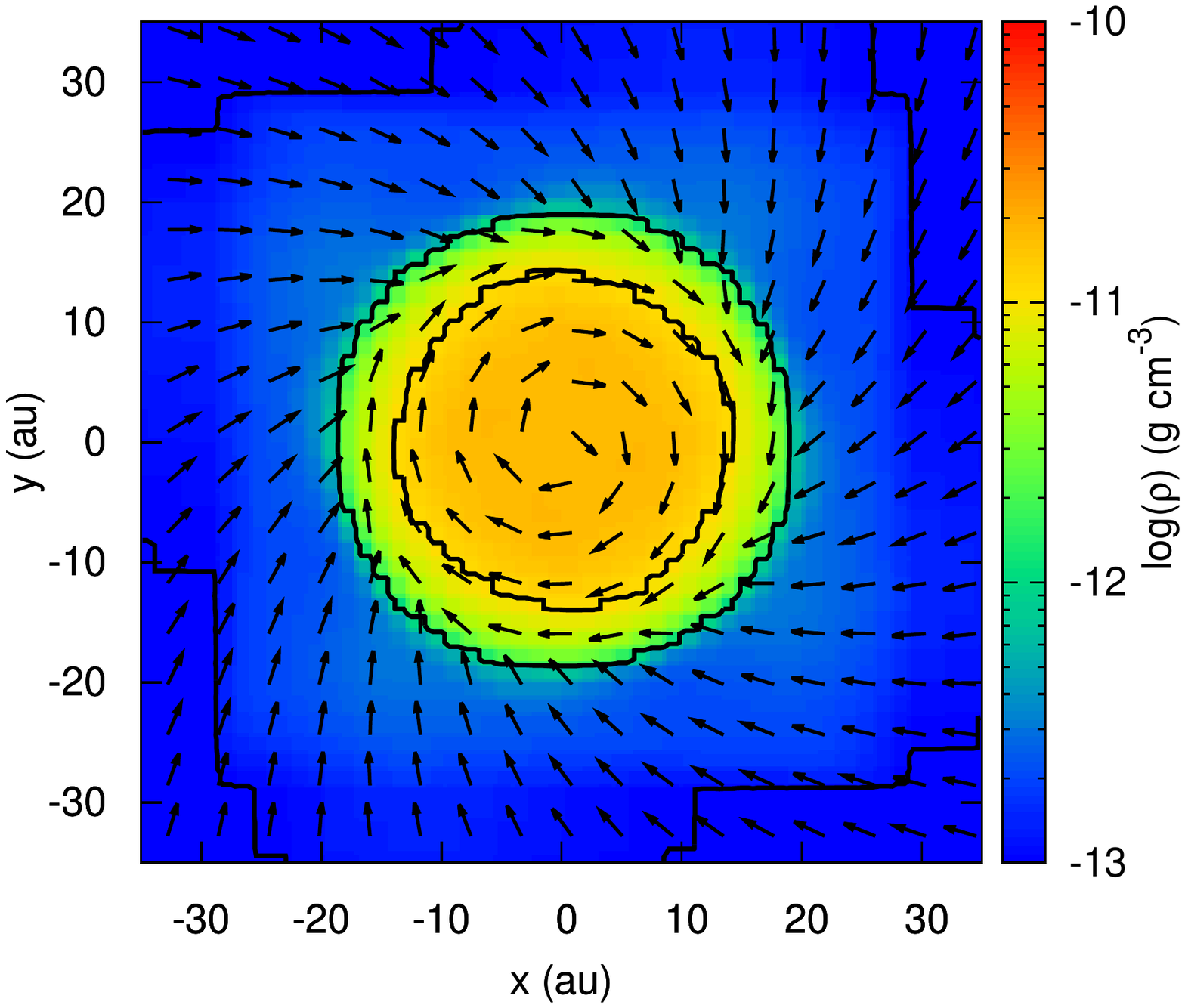}
  \caption{Density slices of the mid plane for (from left to right) Th0, NoHall, and Th180. Black vectors represent the direction of the gas velocity.}
  \label{FigSlicerho}
\end{center}
\end{figure*}

In the following simulations, we use the same initial conditions as \citet{Tsukamoto2015} (hereafter T15). The purpose is to assess the validity of our numerical methods in realistic conditions by comparison with an independent implementation.
The initial cloud is a uniform sphere of $M=1$ M$_\odot$, with a radius $R=2957$ au and a temperature of 10 K (thermal-to-gravitational energy ratio of $\alpha=0.3$). The sphere undergoes a solid rotation characterized by a rotational-to-gravitational energy ratio of $\beta=0.011$. The initial magnetic field is uniform, either parallel ($\theta=0\degree$) or anti-parallel ($\theta=180\degree$) to the rotation axis, with a mass-to-flux ratio of $\mu=4$. All three non-ideal MHD effects are included. We also perform another simulation without the Hall effect for comparison. The three cases are summarized in table \ref{TableSimu}. Contrary to T15, we do not include the effects of radiation-hydrodynamics, and instead use the barotropic equation of state \eqref{eqeos} to compute the temperature. While the first core is not able to do its second collapse with such EOS, the temperature rise prevents high densities to be reached quickly, which would slow the simulation. We are here interested in the formation of structures in and around the first core and the disk rather than the formation of a protostar. We define the disk as a rotationally-supported structure, with the same criteria as in \citet{joos} and paper 1.
Contrary to T15, the gas outside the sphere is at rest, with a density 30 times lower than the sphere, and boundary conditions are periodic. The refinement criterion is 8 points per Jeans length.

\begin{table}
  \caption{List of simulations}
\label{TableSimu}
\centering
\begin{tabular}{lll}
\hline\hline
  Name & Hall effect & Angle \\
\hline
  Th0    &  Yes &  $\theta=0\degree$  \\
  Th180  &  Yes &  $\theta=180\degree$  \\
  NoHall &  No  &  (Direction irrelevant) \\
\hline
\end{tabular} 
\end{table}

Simulations have been performed with the generalized moncen slope limiter with a coefficient 1.05 for the magnetic field and minmod for the other variables. We use the shallow slope limiters to prevent any overshooting of magnetic field while reconstructing states at cell interfaces. In the following sections, $t_\mathrm{C}$ denotes the formation time of the first Larson core, i.e. when the maximum density reaches $10^{-13}$ g cm$^{-3}$.

\subsection{Magnetic resistivities}

To compute the resistivities of non-ideal MHD effects, we use the table of \citet{2016A&A...592A..18M}, which contains the equilibrium abundances of a reduced chemical network across a wide range of density and temperature. The network includes species relevant to the star formation environment and grains following the MRN size distribution \citep{mathis}. They take into account thermal ionisations, the thermionic emission of grains \citep{deschturner} and the grain evaporation. During simulations, the non-ideal MHD resistivities are calculated for each cell using the local state variables.

\subsection{Results}

We choose to stop the slower simulation, Th0, at $t=t_\mathrm{f}= t_\mathrm{c} + 1400$ years, and the two faster simulations at $t= t_\mathrm{c} + 3000$ years. At the 1400 years mark, the maximum density is $\rho_\mathrm{max}=10^{-10}$ \gcc for Th0, $\rho_\mathrm{max}=10^{-11}$ \gcc for Th180 and $\rho_\mathrm{max}= 6.6 \times 10^{-11}$ \gcc for NoHall. 
Figure \ref{FigSlicerho} displays density maps of the mid-plane for the three simulations at this time. As in T15, the Hall effect reduces the magnetic braking in the anti-parallel case, allowing the formation of a large disk, and enhances it in the parallel case, speeding up the collapse. The Hall effect modifies the size of the disk by up to 50\% in this setup, resulting in a factor 2 between the models with parallel and anti-parallel magnetic fields. The disks in Th180 and Nohall develop a $m=2$ instability at $t=t_\mathrm{c} + 2000$ years and $t=t_\mathrm{c} + 3000$ years, respectively.

\begin{figure}
\begin{center}
\includegraphics[trim=7cm 2cm 5cm 3cm, width=0.45\textwidth]{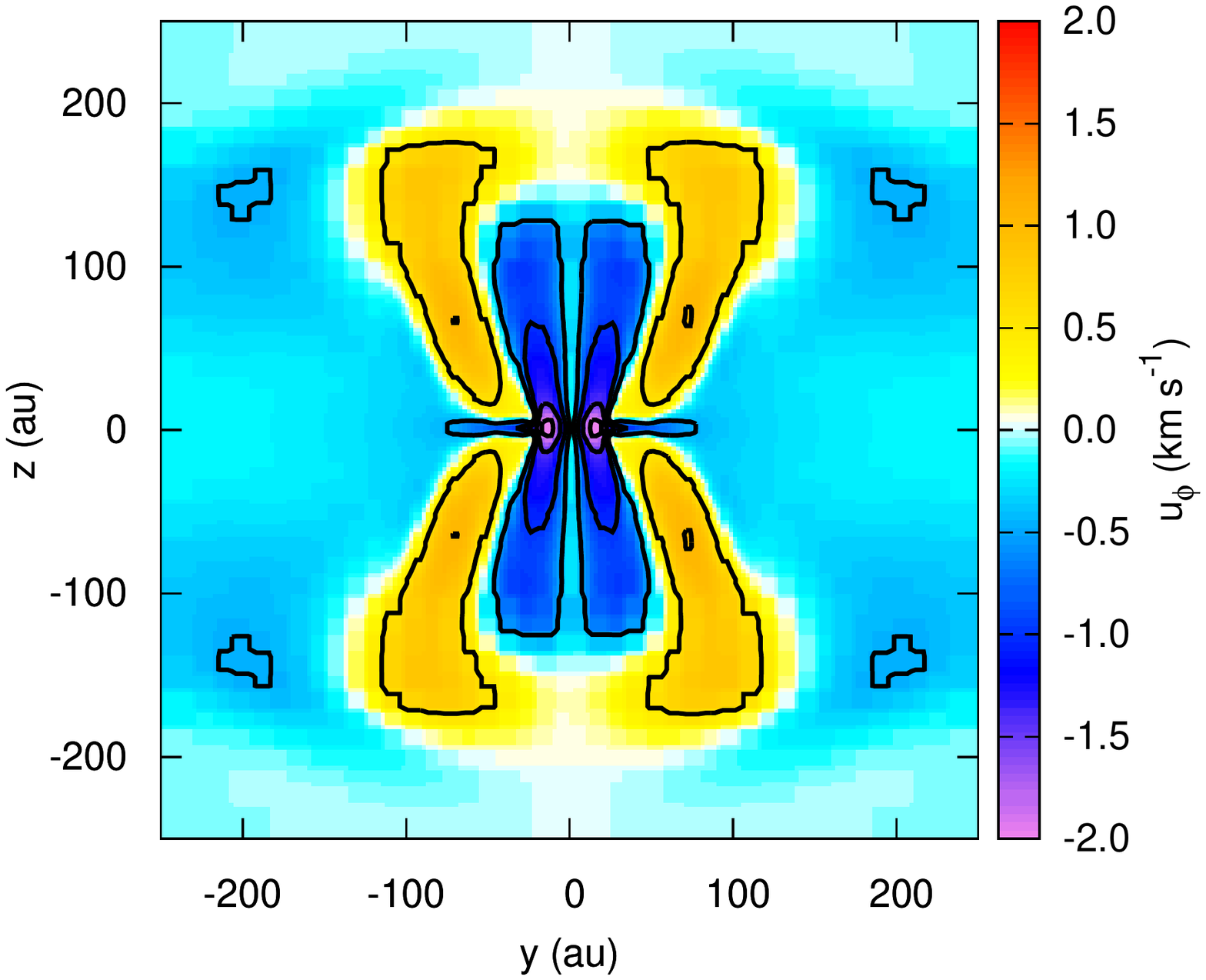}
\includegraphics[trim=7cm 2cm 5cm 3cm, width=0.45\textwidth]{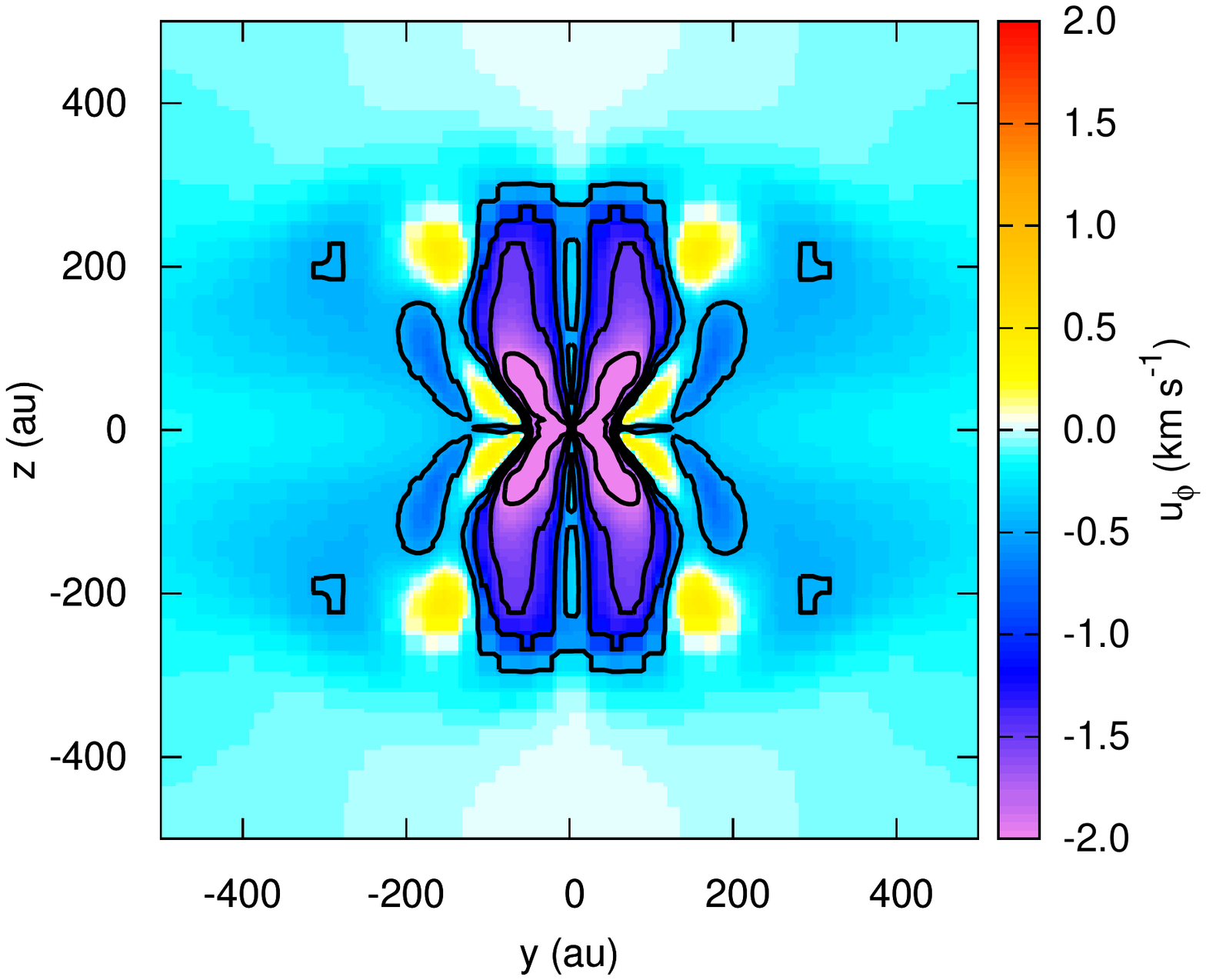}
  \caption{Azimuthal velocity slice in the plane $y=0$ (edge-on) for Th180. Blue-purple colors represent the rotation in the negative direction, and red-yellow colors the rotation in the positive direction. Top panel : $t-t_\mathrm{C}=1.4$ kyr, bottom panel : $t-t_\mathrm{C}=2.9$ kyr.}
  \label{FigSlicevtheta}
\end{center}
\end{figure}

Figure \ref{FigSlicevtheta} represents the azimuthal velocity maps with the disk seen from the edge for Th180, as figure 5 of T15, at $t=t_\mathrm{C}+1400$ year and $t=t_\mathrm{C}+300$ year. The global rotation (in the negative direction in this simulation) is reinforced in the mid plane by the Hall effect, leading to a high rotation velocity, up to $u_\phi \approx 2$ km s$^{-1}$. Counter-rotating envelopes form in this case, with velocity of $u_\phi \approx 1$ km s$^{-1}$. The scale and geometry of the various rotating regions match almost perfectly the figure 5 of T15. This geometry evolves over time. The infalling negatively-rotating gas eventually mixes with the positively-rotating envelopes, that almost disappear at the end of the simulation, $1.5$ years after this snapshot (see bottom panel of figure \ref{FigSlicevtheta}), for a total lifetime of $\lesssim 3000$ years in this case.

\begin{figure}
\begin{center}
\includegraphics[trim=3cm 2cm 3cm 2cm, width=0.45\textwidth]{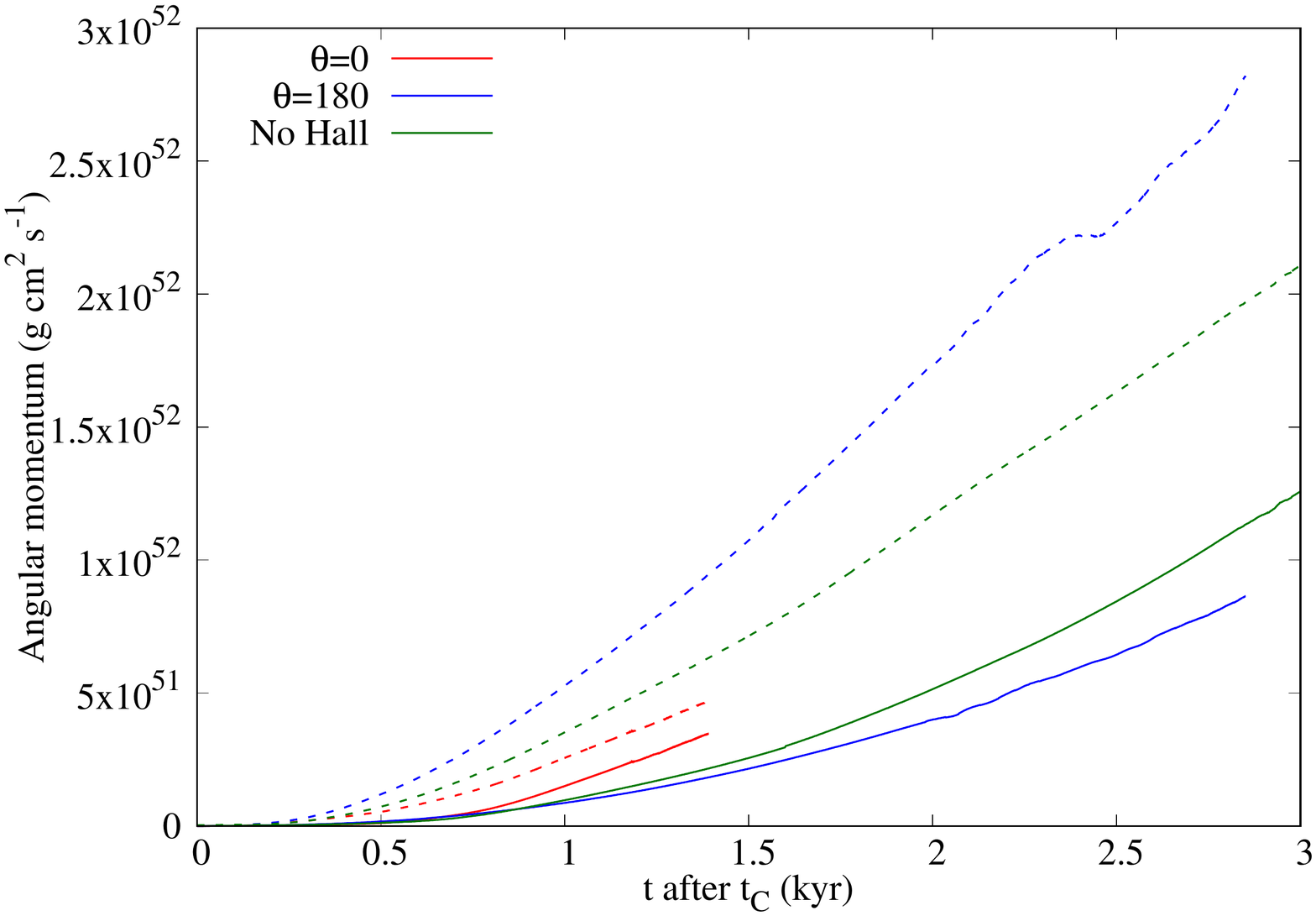}
  \caption{Angular momentum evolution after the first core formation for Th0 (red curves), Th180 (blue curves) and Nohall (green curves). The solid lines represent the "excess" of angular momentum in the whole box compared to $t=t_\mathrm{C}$.}
  \label{FigAngularmomentum}
\end{center}
\end{figure}

We now compare the angular momentum in the various cases. Figure \ref{FigAngularmomentum} shows the angular momentum contained in the core ($\rho > 10^{-13}$ \gcc) and the disk. Unsurprisingly, the run in which the Hall effect enhances (reduces) the magnetic braking has the lowest (highest) angular momentum, with a factor two between both, and NoHall being an intermediate case.
The solid line is the total angular momentum in the simulation box, shifted such as it equals zero at $t=t_\mathrm{c}$. It then represents the accumulated error on the total angular momentum as function of the time after the first core formation. In all three cases, this quantity increases at similar rates. After 1400 years, the excess reaches $\approx 2-3 \times 10^{51}$ \gcms, which is less than 1\% of the total angular momentum $L_\mathrm{tot} \approx 3.7 \times 10^{53}$ \gcms. At $t=t_\mathrm{c}+3000$ year, this error represents $\approx 2.7$\% of the total angular momentum. 
 The relative value between the dashed and solid lines gives the upper-limit of the fraction of the disk's angular momentum that can be attributed to the spurious error linked to the Hall effect. This fraction is 30\% in Th180 and 80\% in Th0. However, given that the "excess" of angular momentum in NoHall is similar to Th0 and Th180 (and even higher in this case), the spurious fraction due to the Hall effect is most likely minor.

\section{Discussions} \label{SecDiscussions}

In the simulations with initial rotation, the angular momentum excess due to the Hall effect is now a minor factor with respect to the total angular momentum, but our results show that conservation is still not perfect, independently of the Hall effect. This can however be significantly improved by simply increasing the resolution to 16 points per Jeans length, which remains computationally affordable. Also, the additional thermal support provided by the stiff EOS keeps the gas away from the center. Any numerical error on the linear momentum therefore results in a larger error on the angular momentum compared to a case with a more compact core.

A quantitative comparison to T15 is difficult here because we use a stiff barotropic EOS instead of radiation-hydrodynamics, and our chemical models to compute the resistivities are different. We also did not evolve the simulation for such a long time. Though, we reproduce very well their results, especially concerning the size of the disk and the geometry of the various rotating regions.

Counter-rotating envelopes of 200 au scale develop on each side of the disk, but eventually disappear in $3$ kyr. While such structures could be observed by modern instruments, as was claimed by \citet{2018ApJ...865...51T}, their short lifetime makes their discovery extremely unlikely. Different magnetic field inclinations can produce more prominent envelopes, especially $\theta=90\degree$ and $\theta=135\degree$ \citep{2017PASJ...69...95T}. We however do not expect them to live significantly longer. Should one be detected, we would expect to find it in the vicinity of a first hydrostatic core.

\section{Conclusions} \label{SecConclusions}

The angular momentum conservation has clearly been improved by the minor modification of our numerical scheme. While the issue has not completely disappeared, the spurious generation of angular momentum has been reduced by one order of magnitude. It is now negligible in regard to the disk's angular momentum of a simulation with initial rotation and with realistic resistivities.
The origin of the problem is still unknown, but we can now be confident in the results of our simulations with the Hall effect if we have sufficiently high resolution.

\begin{acknowledgements}
  We greatly thank Geoffroy Lesur and Yusuke Tsukamoto for the interesting ideas and discussions. We also thank the anonymous referee for his thorough report and comments which helped to improve the clarity of the manuscript. We acknowledge financial support from an International Research Fellowship of the Japan Society for the Promotion of Science and from the "Programme National de Physique Stellaire" (PNPS) of CNRS/INSU, CEA and CNES, France. Computations were performed at the Common Computing Facility (CCF) of the LABEX Lyon Institute of Origins (ANR-10-LABX-0066).
\end{acknowledgements}

\begin{appendix}

  \section{Scheme convergence}\label{AppConvergence}

\begin{figure}
\begin{center}
\includegraphics[trim=3cm 1cm 3cm 2cm, width=0.45\textwidth]{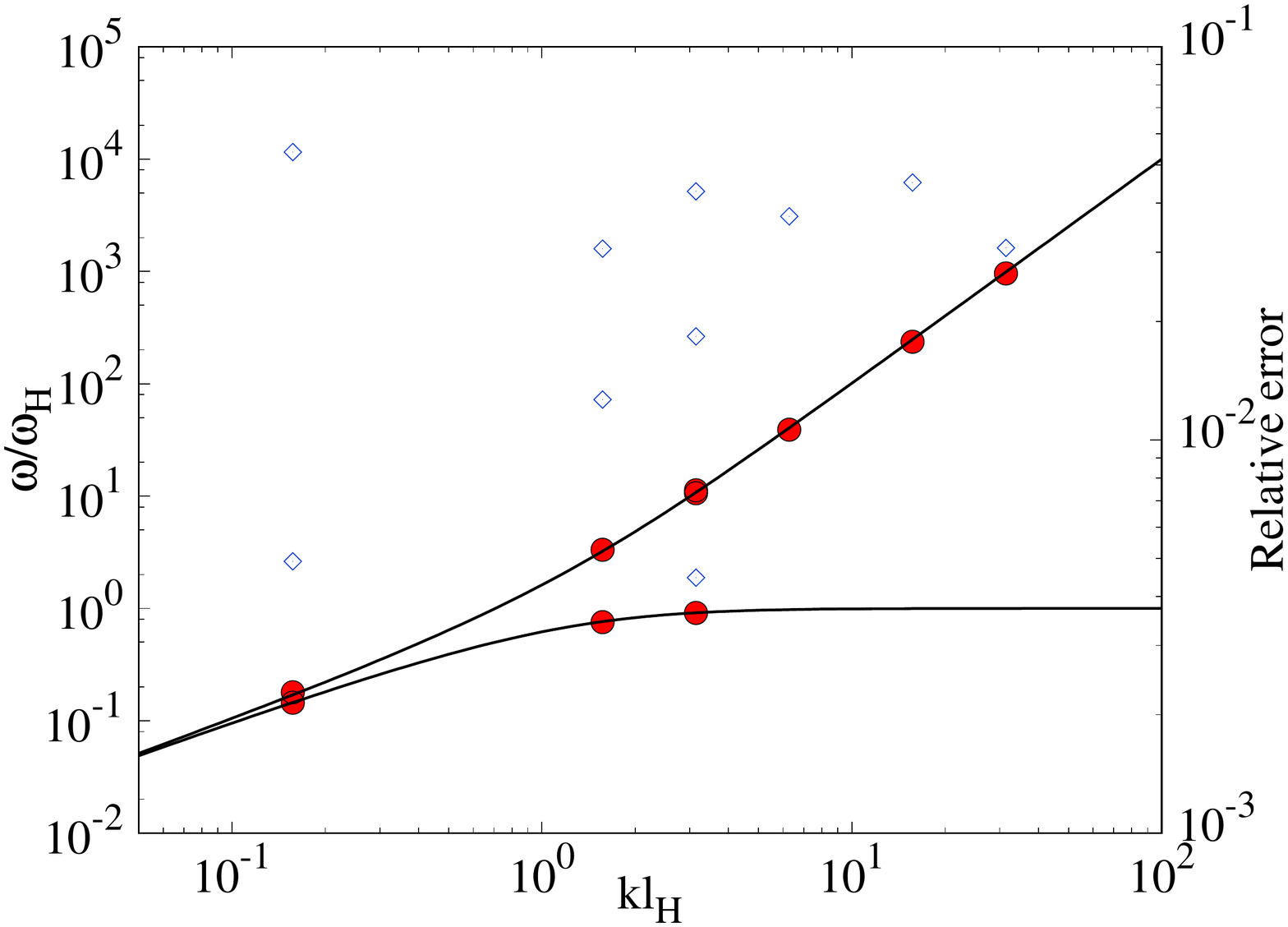}
\includegraphics[trim=3cm 1cm 3cm 1cm, width=0.45\textwidth]{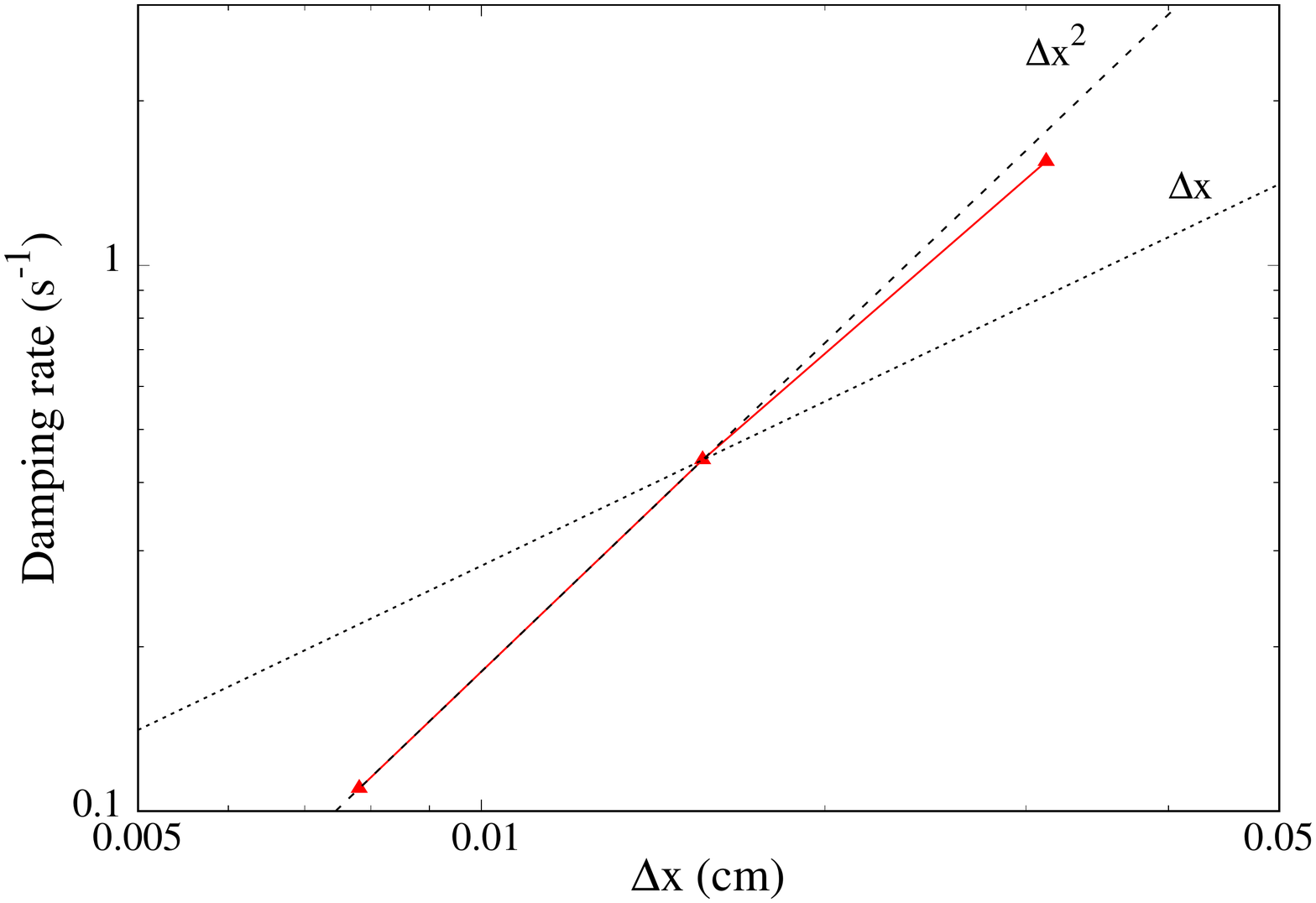}
  \caption{Top panel : Dispersion relation of the numerical scheme. Red points are the measured oscillation frequencies normalised by $\omega_\mathrm{H}=c_\mathrm{A}^2/\eta_\mathrm{H}$ as a function of $kl_\mathrm{H}$ with $l_\mathrm{H}=\eta_\mathrm{H}/c_\mathrm{A}$. Solid black lines represent the theoretical dispersion relation, while blue points show the relative error of each measure. Bottom panel : Damping rate of whistler waves as a function of resolution, compared with $\Delta x$ and $\Delta x^2$ scaling.}
  \label{FigReldisp}
\end{center}
\end{figure}

  Our new scheme successfully passes all tests described in paper I. We show here the tests described in section 4.1 \citep{2002ApJ...570..314S}.
  In a periodic box with a uniform density and pressure, we setup a magnetic wave $\mathbf{B}=B_x \mathbf{e}_x + B_y(x) \mathbf{e}_z$, where $B_x=0.1$ G and $B_y(x)=10^{-3}\cos(2\pi k x)$ G, with $k$ the wave number (and the number of periods in the box). Only the Hall effect is present with a constant resistivity $\eta_\mathrm{H}$. First, we use $k$ and $\eta_\mathrm{H}$ as parameters to find the dispersion relation as in section 4.1.1 of paper I. $k$ ranges from 5 to 20, and $\eta_\mathrm{H}$ ranges from $5 \times 10^{-3}$ to $0.1$ cm$^2$ s$^{-1}$. The resolution of the box is uniform with $128^3$ cells. We use the same slope limiters as in section \ref{SecRot}, i.e. generalized moncen with a coefficient 1.05 for the magnetic field, and minmod for the other hydrodynamic variables.
The top panel of figure \ref{FigReldisp} represents the results, with the normalized frequency of propagation $\omega / \omega_\mathrm{H}$, $\omega_\mathrm{H}=c_\mathrm{A}^2/\eta_\mathrm{H}$ as a function of $kl_\mathrm{H}$, with $l_\mathrm{H}=\eta_\mathrm{H}/c_\mathrm{A}$. We recover the dispersion relation of the Hall effect (equation 18 of paper I) with an error of less than 5\% for most points.
  For the convergence test, we fix $k=5$, $\eta_\mathrm{H}=0.005$ cm$^2$ s$^{-1}$ and we use resolutions of $32^3$, $64^3$ and $128^3$. While propagating, the wave is dissipated by the numerical diffusion (after ~5 periods for the low resolution case), and we compute the damping rate for each case, as displayed in the bottom panel of Figure \ref{FigReldisp}. It shows that the new method keeps the second order spatial convergence of the numerical damping.

\end{appendix}

\bibliographystyle{aa}
\bibliography{MaBiblio}

\end{document}